\documentclass[a4paper]{article}
\usepackage{amsmath,amssymb,bm,cellspace,graphicx,rotating}

\pdfoutput=1
\newcommand{\dd}{\mathrm{d}}
 
\begin{document}
\title{Automatic
  generation of non-uniform random variates for arbitrary pointwise
  computable probability densities by tiling}
\author{\normalsize \textsc{Daniel Fulger}$^{1,2}$
and \textsc{Guido Germano}$^1$
\smallskip \\ 
\normalsize $^1$ Computer Simulation Group, FB 15\\
\normalsize Philipps-University Marburg, 35032 Marburg, Germany\\
\normalsize www.staff.uni-marburg.de/$\sim$germano
\smallskip \\ 
\normalsize $^2$ Complex Systems Lagrange Laboratory\\
\normalsize Institute for Scientific Interchange\\
\normalsize Viale Settimio Severo 65, 10133 Torino, Italy}
\maketitle
\begin{abstract} We present a rejection method based on recursive
  covering of the probability density function with equal tiles.  The
  concept works for any probability density function that is pointwise
  computable or representable by tabular data. By the implicit
  construction of piecewise constant majorizing and minorizing functions
  that are arbitrarily close to the density function the
  production of random variates is arbitrarily independent of the computation
  of the density function and extremely fast. The method works unattended for
  probability densities with discontinuities (jumps and poles).
  The setup time is short, marginally independent of
  the shape of the probability density and linear in table
  size. Recently formulated requirements to a general and automatic
  non-uniform random number generator are topped. We give benchmarks together
  with a similar rejection method and with a transformation method.
\end{abstract}

\section{Introduction and background}\label{sec:intro}

This article introduces a setup method for a rejection algorithm
for the production of random numbers with an arbitrary probability
density functions (PDF) with finite support and that is at least
pointwise computable.  The key feature is fast production of random
variates in computational applications and simple applicability
to any probability density with any number of modes.  The
principle consists of covering the surface under the PDF with
\emph{equal} tiles for which no a priori information is required.  The
speed of random number production is arbitrarily independent of the
shape and computational cost to evaluate the PDF.
Prior to the introduction of the method we give a brief review of the
subject, some existing methods and terminology.

The two topics of uniform and non-uniform random number generators (RNG) are
rather disjoint, with little overlap in the literature and the associated
communities. This is not surprising as the respective problems are quite
distinct and can be considered as subsequent tasks. A non-uniform RNG requires
a uniform one, usually employed as a black box, and the quality of the former
depends on the quality of the latter. Contrary to what one might have expected,
the past 15 years have seen a considerable development of uniform RNGs (URNGs).
For almost half a century since the beginning of the age of computing in the
1950s, uniform random numbers have been produced with linear congruential
generators, which are based on the recurrence relation $I_{j+1} = a I_j + c
(\mod m)$. After shortcomings began to be noticed in the 1960s, much effort
went into reducing them by tuning the parameters, especially $a$ and $m$.
However it took decades until alternative algorithms were discovered, and then
over a dozen appeared within a short time, with the possibility of improving
the quality by combining different methods: the XOR shift URNG
\cite{Marsaglia2003}, the multiply with carry URNG, the linear feedback shift
register URNG, etc. For reviews, see Refs.~\cite{Devroye1989,Knuth1997,
LEcuyer1997,LEcuyer2004,Press2007}.

Fast generation of non-uniform random numbers is important in e.g.\
Monte Carlo simulations~\cite{Ahrens1993, Devroye1989, Leydold2000a,
Hoermann2004, LEcuyer2004, Leydold2003b, Leydold2003c, Marsaglia1984}.
Statistical theory shows how one can produce random variates for any meaningful
distribution. Nevertheless, intelligent mathematical but also purely
computational methodology was developed to achieve speed for
well known analytical and invertible distributions, for non-analytic
but transformable distributions and also for empirical PDFs that only
exist as tabular data. In the earlier days of computing any progress
was taken very
seriously~\cite{Marsaglia1961,Marsaglia1964aa,Marsaglia1964b,Marsaglia1964a}
in applied mathematics but also more recently new perspectives on
seemingly converged methodology on, for example, Gaussian distributed
random numbers can be found~\cite{Leva1992,Rubin2006,Thomas2007b}.

A plethora of mathematically involved publications was inspired by the
practitioner's need to increase the speed and quality of non-uniform
random number production in applications of statistical computing.
Another big driving force is simplicity of application. Special
requirements of initialisation for example are a nuisance.  Each
context and application provides different and often opposite
challenges.  For example the famous Ziggurat method by
Marsaglia~\cite{Marsaglia1984} and its implementation by Marsaglia and
Tsang~\cite{Marsaglia2000} is a non-truncating method within the
narrow class of symmetric, strictly decreasing, analytic and
invertible densities.  Other algorithms existed long before, but the
method's appeal is that the specific \emph{implementation} is even
faster than any other that is specialized entirely on exponential or
normal distributed numbers.  The required initial data structure setup
depends on parameters that so far have been published only for the
exponential and normal distributions~\cite{Marsaglia2000} and are
difficult to derive automatically~\cite{Rubin2006}.  A more
general approach is taken by Ahrens~\cite{Ahrens1993}. This well-known
method is able to process any tabular data that fulfills few
restrictions on smoothness, but the setup and production of random
numbers is slower. The latter two methods are related to more general
strip or slice methods --- already existing for a long
time~\cite{Devroye1989,Marsaglia1964b,Marsaglia1964a,Neal2003,Pang2001}
--- but are of much higher importance in computational applications.
It should be kept in mind that information on the speed of a method is
only meaningful with respect to a particular implementation and
hardware. Some famous methods are actually specialized implementations
that rely on the cache memory of contemporary processors.

There are many collections of specialized non-uniform random number
generators, usually more than one for a particular class of PDFs, each
of them consisting of tailored code.  The generators can be
categorized into two classes: a) A setup of some data structure is
carried out before the first random variate is drawn, and b) a setup
is not needed, e.g.\ with PDFs for which inversion methods
exist.  Clearly, any ``universal'' method will need a setup, as explained
well for instance in Ref.~\cite{Leydold2003a}. In statistical
computing the user taps from these collections of software choosing a
particular generator. Ideally one can also choose gradually between
large setup time and fast generation of random numbers or fast setup
and slow generation. The drawbacks of such collections can be huge
codes, each bug prone, and the increasingly intractable specialties of
the requirements for the setup. Furthermore, the classes of available
distributions are limited in the end. The ultimate goal is a
universal, easy to use and also fast black box
generator~\cite{Leydold2003a}. The definition and limitations of a
\emph{universal} random number generator, however, is often imprecise
in many publications. For example, it does not have any built-in knowledge
of the PDF, except that a meaningful PDF can be known only in
as much as the number of modes is not
infinite~\cite{Ahrens1995} and that the modes are not infinitely thin,
i.e.\ meaningless delta-like functions. For universal applicability
all transformation methods drop out because they cannot be found automatically by a general
algorithm.  Therefore, in the general case only some approximation like pointwise data will be
available to represent the PDF, implying that the support of the distribution
cannot be infinite either.  Any computational approach subdues to this
restriction.  If the PDF can only be evaluated at horrendous costs and
no inversion method exists, then no procedure will be able to produce
usefully random numbers.  Thus, we can assume that the evaluation cost
of the PDF for the appropriate number of points is within a similar
order of magnitude as the overall task within which the random numbers
are used, e.g.\ a Monte Carlo calculation.

In some cases of PDFs with infinite support truncation of the
probability density can be justified with statistical negligibility of
the tails. In few cases of distributions with infinite support and
where the PDF is at best pointwise computable as with the L\'evy
distribution~\cite{Nolan1997,Nolan1999}, which we use as a benchmark,
transformation methods that do sample the infinite range were
found~\cite{Chambers1976}, at least within overflow
limitations.  Such heavy-tailed distributions deserve attention if
truncated early unless justified (or required) by the
application~\cite{Cartea2007a,Cartea2007b,Mantegna1994}; however,
this aspect is largely left undiscussed in the literature.
State of the art methods, e.g.\ Refs.~\cite{Devroye1989,
Leydold2000a,Hoermann2004, LEcuyer2004,Leydold2003b, Leydold2003c},
construct in the setup phase a majorizing (or envelope or comparison) function
and usually also a minorizing (or squeeze) function (see Sec.~\ref{sec:squeeze}
for an introduction) with secants or other
segmentations to be used within a rejection technique with look-up
tables.  In addition to truncation, in many methods it is often also
required to know the approximate or even exact location of the mode
(of which mostly only one is allowed), while the method is still
declared to be suitable for ``arbitrary
PDFs''~\cite{Evans1998,Leydold2000b}.  More limitations to ``general
methods'' are explained in Ref.~\cite{Evans1998}, where it is argued
that the modes of the PDF must be known beforehand for certain
techniques to be suitable for ``arbitrary densities''. The same
authors also construct ``general algorithms'' that depend on concavity
properties and analyticity of the density. It is common in several
methods, e.g.\ adaptive procedures for log-concave distributions to
use the value of maximum density explicitly in the setup of the
approximation of the comparison function~\cite{Devroye1989}.  Another
example is the transformed density method~\cite{Leydold2000}, that
employs a strictly monotonically increasing differentiable transform
such that the transformed PDF is concave.  This method is also
considered universal. In the improved ratio of uniforms
method~\cite{Leydold2000b} the setup is restricted to certain classes
of distributions if the required transform of variables must yield
a region that can be sampled efficiently. In some cases one often
resorts to a rejection technique and the concept of squeeze functions, as also
employed in this paper, to improve the situation.
Yet another general and adaptive approach constructs
a polynomial approximation of the inverse distribution function in
$X=F^{-1}(U)$ that is stored in tables to be used for interpolation
during production~\cite{Hoermann2003}.  For the class of log-concave
distributions Ref.~\cite{Gilks1992} introduces piecewise
exponentials for the approximation of the majorizing and minorizing
functions using previously sampled points of the density function.
This method however is dedicated to the context of Gibbs sampling
where each variate is usually drawn from different densities.
Finally, closing the topic of piecewise approximation, we mention the
approximation of arbitrary densities via a mixture of simpler
densities.  An interesting example is the triangular approximation
giving a piecewise linear approximation of the target density via
many overlapping triangle densities, which recently was also implemented
in hardware~\cite{Thomas2007}.
The most generally applicable method published so far that is also
fast is due to Ahrens \cite{Ahrens1993, Ahrens1995}. It can
deal with more than one mode whithout a priori information
on the modes.

Overall we find inevitably that methods titled ``universal'',
``automatic'', ``black box'', ``out of the box'' and combinations thereof
clearly cannot sample an infinite support of the PDF, are restricted
to certain classes of density functions or are not automatic out of
the box.
But this verdict is much less restrictive in realistic applications of
statistical computing where the requested distributions can always be
represented for example in terms of (interpolated) pointwise data or
other approximations to any computationally sensible accuracy and
finite support limits.  We stress that approximations of the
density function via previous sampling of points or the approximation
of the inverse distribution function $F^{-1}$ as mentioned above are
like other similar concepts a common approach in the field of non-uniform
random numbers~\cite{Devroye1989,Hoermann2003}.  It is accepted to
truncate the originally infinite support, if statistically
justified. Moreover, if the PDF is given as an arbitrarily accurate
approximation, the location of extrema can always be
determined within any required accuracy in finite time.  Our method
is applicable in this context and belongs to the type of rejection and
segmentation methods as the ones by Ahrens~\cite{Ahrens1993} and
Marsaglia~\cite{Marsaglia2000}. In this context and for our claim we
can therefore continue to speak of arbitrary PDFs since this
nomenclature is widely accepted within the literature. Therefore,
in the quest for a universal random number generator, a method can be
called universal if it can process \emph{arbitrary finite density
function data} without any further information. Techniques that take
finite samples of the desired distribution and then try to match this
distribution~\cite{Leydold2000b} are not discussed here.

Lately easy applicability has become more important. The initial motivation
for this work was to overcome the setup difficulty of the Ziggurat method for
the general symmetric monotonic decreasing case. Eventually, we developed a
simpler method for a significantly more general class of PDFs, at the cost of a
moderate performance penalty due to larger memory requirement as compared to the
original Ziggurat implementation of Ref.~\cite{Marsaglia2000} and to specialized
transformation methods, were available.

With the intention to provide a quintessence of recent research and
demands of statistical computing a wish-list of requirements to a
universal random number generator was presented in
Ref.~\cite{Leydold2003a}, which we quote here:

\begin{enumerate}\label{listofreq}
\item Only one piece of code, debugged only once.
\item By a simple parameter choose between fast setup and slow
  generation or long setup time and fast generation.
\item It can sample from truncated distributions.
\item The rejection rate can be made as close to zero as desired,
  i.e.\ as close to inversion as one wants.
\item The setup time is independent of the density function and is
  faster than many specialized generators.
\item The quality of the non-uniform random numbers is as good as the
  underlying uniform random numbers.
\end{enumerate}

Point 5 should be made more precise. It refers to the independence of
the shape of the density function, but a density with complicated
shape usually requires more information, in particular in regions of
high curvature.  If the input size increases, the setup time is indeed
allowed to grow.  There is no obvious universal measure that gives the
minimum input size required for the suitable representation of a
function. This is responsibility of the scientist.  Several of the
methods mentioned in the above overview are considered to meet these
requirements. This is also the case with the method presented here
plus additional relaxations with respect to the properties of the PDF.

In Sec.~\ref{sec:tiling} the tiling is introduced along with some numerical
considerations, an explanation of the role of the squeeze
function, and a proof of correctness.  Sec.~\ref{sec:discont} shows
that the tiling procedure is capable of dealing \emph{unattended} with
poles and other discontinuities.
Sec.~\ref{sec:measurements} gives benchmarks of typical and
a-typical situations.
Sec.~\ref{sec:conclusion} summarizes and provides a short discussion. 
We chose for comparison well-known methods for
specialized distributions (Gaussian and exponential) but also a
difficult non-analytic distribution (Appendix) for which a transformation
method is available.  We explain computational issues
that are usually ignored but are decisive for performance.
This serves in placing the tiling method into the right context among other
methods with respect to speed and applicability.

\section{The tiling and numerical considerations}\label{sec:tiling}

\subsection{The tiling procedure}\label{subsec:tiling}

For any computational task a PDF with finite support, even one with a non-invertible
distribution for which no specialized method exists, can be represented
either as a) a sufficiently good approximation that can be evaluated
sufficiently fast, e.g.\ by series expansion or polynomials, or
b) as tabular data for interpolation. Any
feature of a meaningful PDF $f(x)$ can be represented in the latter
case by varying the sampling density of the tabular data that
represents the PDF in the form $(x_1,f(x_1)), (x_2,f(x_2)),\dots$
\cite{Ahrens1993}. However, since
the tiling is completely independent of such considerations, we will
simply speak of ``evaluating $f(x)$''.  Furthermore, it can safely
be assumed that the PDF can be evaluated in a finite time comparable to the
duration of the application within which the random variates are to be
used.  With these prerequisites the determination of local extrema
is achievable in $\mathcal{O}(N)$ where $N$ is the number of data points.

\begin{figure}[tb]
   \begin{center}
   \parbox{0.7\textwidth}{\input{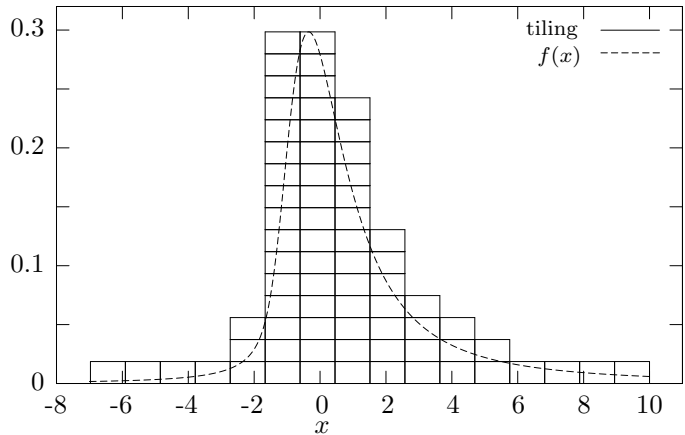}}
   \end{center}
        \caption[Tiling of an asymmetric density function] {\label{fig:asymmetric}
          For the intuitive introduction of the tiling procedure the
          plot shows an early refinement stage in the tiling of a truncated
          asymmetric L\'evy PDF with parameters used as an arbitrary example.}
\end{figure}

\begin{figure}[tb]
    \input{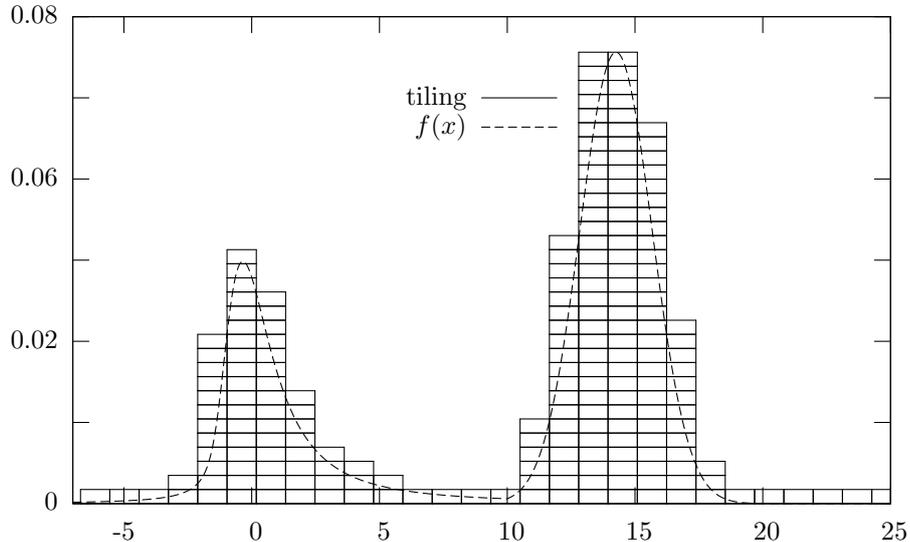}
    \caption[Tiling of a bimodal probability density function]{\label{fig:bimodal} 
    Tiling of a bimodal probability density function.
    $f(x)$ is composed of two L\'evy density functions with parameters
    $\alpha=1,\ \beta=0.7,\ \gamma=1$ (left part) and $\alpha=2,\ \beta=1,\ \gamma=1$ (right
    part); the heights are adjusted to fit the curves seamlessly at
    $x=10$. }
\end{figure}

For the following considerations the integral over the density
function is required only up to a constant factor $C=\int_a^b f(x)
dx$.  The rejection method does not require $C=1$.
The tiling concept is simple, see Fig.~\ref{fig:asymmetric}: The area
under the PDF $f(x),\ x\in[a,b],$ is covered with rectangular tiles
of equal area.  The procedure starts from one single tile $b-a$ wide
and $\max(f(x))$ high. Choosing an initial tile larger than required
by the support and the maximum did not show significant influence on
the outcome in all cases we tested.  The initial tile
is split into four equal tiles, and so on
recursively. At each refinement cycle \emph{all} tiles are
split. Those that lie entirely above the PDF are discarded in each
cycle. The splitting can be stopped once a given accuracy of the
covering is reached; Sec.~\ref{sec:squeeze} explains the details of
the calculation of this condition. Fig.~\ref{fig:asymmetric} shows a
truncated asymmetric L\'evy PDF with parameters $\alpha=1, \beta=0.7,
\gamma=1, \delta=0$ according to the $S_0$-parametrization
convention~\cite{Nolan1997,Nolan1999}.  We use this distribution as an
arbitrary example for comparisons that provides fat tails and for
which a fast transformation method is available; for details see
Sec.~\ref{sec:levy}.  The support is chosen small to produce
deliberately a visible truncation.  Fig.~\ref{fig:bimodal} shows the
tiling of a bimodal PDF.
The above \emph{recursive} procedure may be considered the
most elegant and simple way to construct the tiling. For the subsequent
production stage it is irrelevant however if the tiling was constructed,
for example, by plastering, i.e.\ starting from a small initial tile somewhere
within the support.

Thus, the tiling constructs a piecewise constant majorizing function
$g(x)$ of the PDF $f(x)$, with $g(x) \geq f(x)\ \forall x \in
[a,b]$. The closer $g(x)$ to $f(x)$, the better. The universal von
Neumann rejection method has two main steps:
\begin{enumerate}
\item[a)] Generate a random $X \in [a,b] \sim
g(x)$ and a random uniform $Y \in [0,g(X)]$.
\item[b)] Accept $X$ if $Y<f(X)$, otherwise reject it and repeat the procedure.
\end{enumerate}
The rejection rate is given by the
ratio $R$ of the areas under the PDF and the comparison function:
\begin{equation}\label{eq:rejrate}
R=1-\frac{\int_a^b f(x)dx}{\int_a^b g(x) dx } =1-\frac{1}{N S}\int_a^b f(x)dx.
\end{equation}
The denominators correspond to the sum over all $N$ tile
surfaces $S$ which are equal.

At this point one could think that an adaptive scheme would be more appropriate,
e.g.\ only tiles intersected by the PDF are split or even deformed to fit the boundary better,
or tiles lying below the PDF are merged. Indeed this is common practice in computer graphics and some 
approximation methods. However, this measure to save memory is not recommended here; actually, it is
to be avoided for the sake of simplicity and speed. In the production stage the
probability of random selection of a tile would have to be proportional
to its area to guarantee uniform probing. This is more complicated and slower especially
if the area ratios are not integer. Moreover, a uniform random 
coordinate is more expensive to produce in shapes other than rectangles. Additional details
to why the segmentation into equal areas is crucial are given 
in numerous publications~\cite{Ahrens1993,Devroye1989,Marsaglia1984,Marsaglia2000}.
Instead of using strips of different height and width or even different shape as in other methods,
here we suggest \emph{equal} tiles as the clearly simplest and fastest approach. This is the key
idea in this work.

Now the von Neumann rejection can be implemented with a modified first step:
\begin{enumerate}
\item[a)] Generate a random tile index $i = 1,\dots,N$;
generate a random coordinate $(X,Y)$ within tile $i$.
\item[b)] Accept $X$ if $Y<f(X)$, otherwise reject it and repeat the procedure.
\end{enumerate}
This way we are able to sample efficiently the majorizing function $g(x)$.
Moreover, the evaluation of the condition in b) is hugely sped up using the
implicitly constructed minorizing function as explained in more detail in the following 
Sec.~\ref{sec:squeeze}.\\

Although the sampling with tiles seems sufficiently intuitive and
equivalent to analogous methods of this
kind~\cite{Marsaglia1984,Marsaglia2000} we give nevertheless a
reasoning on the correctness.

\textsc{Theorem} The introduced sampling of the comparison function is
  equivalent to the standard von Neumann sampling, i.e.\ $g(x)$ is
  sampled uniformily within all tiles generated on the support
  $x\in[a,b]$.

\textsc{Proof} Define $I=\{i_1,i_2,i_3,...,i_N\}$ the set of tile
  indices and $I_j\subset I$ the subset of all indices
  $i_1^j, i_2^j,...$ corresponding to a particular tile column $j$
  with width $\Delta x=(b-a)/r$, where $r$ is the number of columns. 
  Thus $\bigcup_j I_j=I$.  Construct an 
  bijective mapping $i_k^j\to n^j_l$ with $n_l < n_{l+1}$. The
  mapping is purely a renaming of indices in column $j$. So we have
  $n^j_l=1,\dots,n^j_\mathrm{max}$, $n^j_\mathrm{max}=g(x)/\Delta y$
  where $\Delta y$ is the height of the tile.  Note that within column
  $j$ the function $g(x)$ is constant. Now define a random number
  $Y^j=n^j_l\ u\ \Delta y$ with uniform random $u\in[0,1)$. The index
  $n^j_l$ is random by the random choice of $i_k^j$ and the subsequent
  mapping.  Then $Y^j\in[0,g(x))$ is a uniform random number in column
  $j$ and we arrive at the standard situation of the rejection method
  for the interval $x\in \Delta x_j$: Generate a uniform coordinate
  $(X^j,Y^j)$ with uniform $X^j\in\Delta x_j$ and reject $X^j$ if
  $Y^j>f(x)$. The sampling of $j$ is implicitly proportional to the
  size of $I_j$, i.e.\ the height of column $j$, due to the uniform sampling of
  tile indices $i\in I$ and $\bigcup_j I_J=I$. Therefore $X^j$ is
  sampled as desired according to $g(x)$ and the sampling of pairs
  $(X,Y)$ is achieved with $X\in\bigcup_j \{X^j\}\sim g(x)$.
\hfill $\Box$

The correctness of the standard rejection method can be taken
for granted since the seminal paper by John von Neumann~\cite{vonNeumann1951}.

\subsection{Implicit squeeze function}\label{sec:squeeze}

The tiling also constructs implicitly a so-called squeeze function
$q(x)$ that fulfills the condition $q(x)\leq f(x) \leq g(x) \leq g(x)$ 
within the required interval $[a,b]$. This is the usual definition of
the squeeze and comparison functions, see for example
Ref.~\cite{LEcuyer2004}. $q(x)$ is the upper edge of the top tiles lying completely 
underneath $f(x)$, or equivalently the bottom edge of the tiles intersected by $f(x)$.  
The role of the squeeze function is to reduce
the number of evaluations of $f(x)$ if $q(x)$ can be evaluated faster:
In the setup all tiles below $f(x)$ are labeled and the test $Y\leq
q(X)$ involves no computation --- just one label look-up. Actually $Y$
must not be generated at all for tiles that are not intersected by $f(x)$.
The latter is the key advantage of the squeeze function.  Thus the
following modified steps implement the von Neumann rejection:
\begin{enumerate}
\item[a)] Generate a random tile with index $i =1,\dots,N$;
generate a random $X$ within tile $i$.
\item[b)] Look up if tile $i$ is labeled as ``$<f(x)$''. If yes, accept $X$.
Otherwise generate $Y$ within tile $i$ and compare $Y<f(X)$. If yes, accept $X$.
Otherwise reject it and repeat the procedure.
\end{enumerate}
With dense tiling most $X$ are accepted in b) by
one table look-up only without the generation of a second real
coordinate $Y$. The PDF itself is hardly ever evaluated.
The relative number of evaluations of $f(x)$ per non-uniform variate
is given by
\begin{equation}
  E=1-\frac{1}{N S} \int_a^b q(x) dx. 
\end{equation}
The integral over the squeeze function is given by the sum of all
tile surfaces not intersected by $f(x)$.
Thus, the number of evaluations of $f(x)$ can be greatly reduced and
is equal to the area fraction of the border tiles.  Both
numbers $R$ and $E$ are cheaply calculated on the fly, so that the
resulting rejection rate can be pre-imposed as a condtion for the tile
refinement.  The latter results will be reconsidered in Sec.~\ref{sec:cutoff}
on the distribution cutoff.

To have a better measure of the ``quality'' of $g(x)$ and $q(x)$ we
estimate an upper limit for the probability density $p_E$ that $f(x)$ must be
evaluated for one non-uniform random number. Define $\Delta
x:=(b-a)/n$ where $n$ is the number of columns, so $\Delta x$ is simply
the final width of the tiles.  For $\Delta x \ll b-a$, i.e $n$
sufficiently high, $f(x)$ can be assumed linear in the interval
$\Delta x$. Then
\begin{equation}
  p_E(x, \Delta x) \propto \frac{b-a}{r}\,\frac{\dd \log f(x)}{\dd x}.
\end{equation}
This expression is deduced from the ratio of areas contained in a tile
column corresponding to $Y\leq q(x)$ and $q(x)<Y \leq g(x)$
respectively.

\subsection{Distribution cutoffs} \label{sec:cutoff}

In the introduction and thereafter we explained that all procedures
that are not specialized to particular analytic and thus invertible
distributions will never sample an infinite support. Considerations on
the appropriate cutoff apply only to special distributions
\cite{Devroye1989}.  If the support of the PDF $f(x)$ is infinite,
a general algorithm will inevitably reduce it to a reasonable finite
interval $x\in[a,b]$. It is the scientist's responsibility to control
appropriately these support limits.
 
However, the period length $L$ of the $[0,1]$-uniform generator used in
the sampling along the abscissa must satisfy the condition $f(x)<1/L$
at both limits $a,b$ \cite{Ahrens1995}. This situation appears
for example in the standard rejection method or the Ziggurat method.
In the latter, the $[0,1]$-uniform generator
must sample the whole bottom strip.  The sampling procedure in
our method lifts this limitation by the number of columns $n = 2^{r-1}$,
where $r =  1, 2,\dots$ is the refinement level: A random integer is generated to sample
a tile and a subsequent uniform $X$ is generated \emph{within} the tile.  In
practically relevant cases the number of tiles will always be
exceedingly smaller than the period length of any sensible random
integer generator. Fat (or somehow long) tailed distributions deserve
attention for the above reason.

\section{Discontinuous probability densities}\label{sec:discont}

The literature also considers density functions which contain
a pole (which numerically is indistinguishable from a cusp)~\cite{Ahrens1995}, i.e.\
$f(x)\to\infty $ as $x\to c^+$ or $x\to c^-$ in the range of interest
$[a,b]$. Within the standard von Neumann rejection
method~\cite{vonNeumann1951} a pole is dealt with as follows: Choose
$\epsilon\ll 1$ and assign the cumulative probability
\begin{equation}\label{eq:Pc}
  P_c=\int_{c-\epsilon}^{c+\epsilon}f(x)dx
\end{equation}
to the interval $[{c-\epsilon},{c+\epsilon}]$, a so-called mass
point. If the [0,1]-uniform deviate is smaller than $P_c$ return
$c$. Otherwise sample from $[a,b]\backslash
[{c-\epsilon},{c+\epsilon}]$. If $c$ and $c \pm \epsilon$ have the
same numerical representation then no better method exists to sample
from $f(x)$. Usually this situation must be treated computationally as
a special case in the setup and production phases.

\begin{figure}[tb]
  \begin{center}
  \includegraphics[height=7.3cm, width=7.8cm]{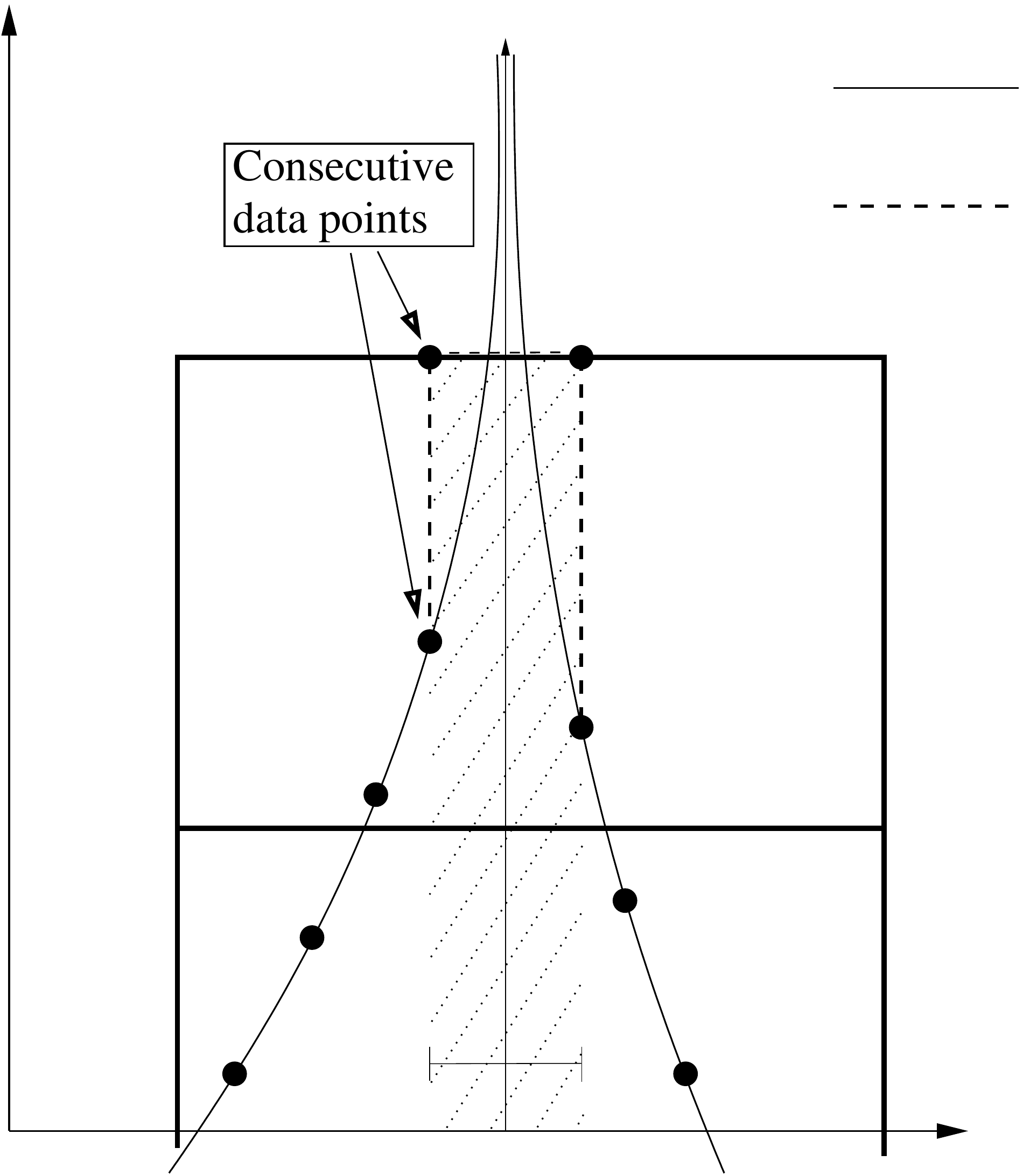}
  \end{center}
  \setlength{\unitlength}{1cm}
  \begin{picture}(0,0)(-6,-0.35) 
    \put(-.3,7.7){ $\infty$}
    \put(1.0,7.1){\small true $f(x)$} 
    \put(0.35,6.35){\small modified $\bar{f}(x)$} 
    \put(3.7,0.5){ $x$}
    \put(-.45,1.15){\small $2\epsilon$}
    \put(-.2,0.5){\small $c$}
    \put(3,5.5){\small $\max(\bar{f}(x))$}
    \put(3,3.0){\begin{sideways}\begin{sideways}\begin{sideways}\small
            Tiling\end{sideways}\end{sideways}\end{sideways}}
\end{picture}
  \caption[PDF containing a pole]{\label{fig:pole} 
    Probability density containing a pole at $x=c$ (schematic).  The
    returned deviates are numerically correct if the modified density
    function $\bar{f}(x)$ fulfills the condition of equal area
    (hatched region) for $x\in[{c-\epsilon},{c+\epsilon}]$ and
    $\epsilon\ll 1$ sufficiently small. The initial tile is chosen
    $\max(\bar{f}(x))$ high. The true or modified PDF can be approximated via data
    points as shown or by any other method that implements this
    condition. The mathematical jumps in the modified PDF can be modeled 
    numerically via two very close consecutive data points along $x$.}
\end{figure}
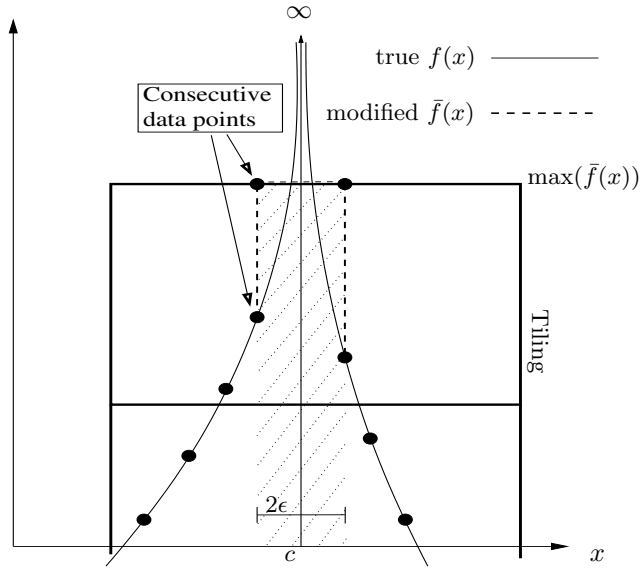

The tiling procedure and subsequent production works unchanged with an
appropriately approximated (or modified) density function as follows.
Fig.~\ref{fig:pole} shows the situation of a density function with a
pole at $x=c$. Figure dimensions, especially the vertical scale, are
exaggerated to convey intuitively the geometry.  Choose
$[c-\epsilon,c+\epsilon]$ and modify $f(x)$ yielding $\bar{f}(x)$ such
that the cumulative probability $P_c$ according to Eq.~(\ref{eq:Pc})
is preserved (hatched area in Fig.~\ref{fig:pole}):
\begin{equation}\label{eq:Pcondition}
  \int^{c+\epsilon}_{c-\epsilon} f(x) dx = \int^{c+\epsilon}_{c-\epsilon}\bar{f}(x) dx 
\end{equation}
which gives the implicit condition for $\max(\bar{f}(x))$:
\begin{equation}\label{eq:maxf}
  \max(\bar{f}(x)) = \frac{1}{2\epsilon}\int^{c+\epsilon}_{c-\epsilon} f(x) dx.
\end{equation}
This is the minimum value of the height of the initial tile. If
$\epsilon$ is chosen sufficiently small with numerical or/and
statistical reasoning, the result will be identical to the procedure
in the standard von Neumann rejection described above.

One has to be aware that the choice of $\epsilon$ as the smallest
representable ``distance'' from the position of the pole is
unnecessarily restrictive.  Any statistical verification requires a
significant number of deviates to fall in the region of the pole to
reveal a possibly too large value for $\epsilon$. 
Depending on the error norm and test
method it is likely to turn out that $\epsilon$ can safely be chosen
magnitudes larger than the initial numerical consideration.  The
statistical needs of the application must be considered in any case.
Thus, there is no generally obvious upper limit for $\epsilon$.

An example application is the scaled
symmetric modified Bessel function of the second kind $ K_0(|x|)/\pi$, 
which is the density of the product $XY$, where $X$ and $Y$ are idependent
normal distributed random numbers. $K_0(|x|)$ diverges at $x=0$.
A possible setting could be the following. 
Restricting the support to $x\in [-15,15]$ accounts for over 99.99999\%
of all mass. With $\epsilon=0.00001$ we get a fraction of $8.03978 \times 10^{-5}$ of 
the mass contained in the interval $[0-\epsilon,0+\epsilon]$.
The number of recursive refinements is given by $\lceil\log_2(30/(2\epsilon))\rceil=21$.
About 235\,000 tiles are retained to cover the density function, 
corresponding to two megabytes of memory in our data format.
Benchmarks for the setup are presented in the next section.
Although it is quicker to directly multiply normal random variates, this example demonstrates
the applicability of the method to other densities with no simple alternative.

PDFs with first order discontinuities or jumps are implicitly
contained in the above case. With a suitable interpolation scheme one
can simply use tabular data to model the jump from one data point
$(x_i,f(x_i))$ to the next $(x_{i+1},f(x_{i+1}))$ and fix $(x_{i+1} -
x_i) \approx \epsilon$ as close as numerically possible. There will be no deviates
falling in $[x_i, x_{i+1}]$. This situation is contained
schematically in Fig.~\ref{fig:pole} as well, showing two consecutive
but very near data points for the left flank of a jump.

\section{Measurements and comparisons}\label{sec:measurements}

In general, speed comparisons are not obvious to do and interpret
and only meaningful with respect to a particular software and hardware
implementation.  With increasing optimization of code, the mathematical
description of a method and its implementation become inseparable. We must
stress this and point to technical aspects that are responsible
for a speed difference of two orders of magnitude, even though the
mathematical/algorithmic description is identical.
The issue of this paper is not pure speed, but portability and easy applicability
in compromise with speed. 

All measurements were performed on a desktop PC with a 2.4 GHz Intel
Pentium 4 processor using the GNU C++ compiler version 3.2.2 on Red Hat Linux.
We explain below the importance of using a multi-tasking operating system in
its standard operation mode with a typical process time slice during the
measurements. This will almost always be the case with applications
in statistical computing.

\subsection{Memory requirements}

At the start the graph of the PDF is embedded in one tile.  The memory
requirements for a Gaussian-like density function and rejection rate
below 0.02~\cite{Marsaglia2000} is never more than a few megabytes.
For details on a uni-modal example as in Fig.~\ref{fig:asymmetric} see
Tab.~\ref{tab:unimodal}, for the bimodal case in
Fig.~\ref{fig:bimodal} see Tab.~\ref{tab:multitable}.  Only
obnoxious density functions with fat tails and many sharp peaks
require more memory. In the tested variations of such extreme
cases using multiple peaks and the support truncated very far out the
memory needed to achieve a rejection rate below 0.02 did not exceed
10 megabytes (about one million tiles). This happens using two numbers
to store the coordinates of one tile and is more than acceptable
for contemporary desktop computers.
We skipped entirely memory optimization and removal of redundancy
since we preferred a clear class structure and
simple data management. Setup time can be reduced via speed optimized
data structures, but that typically increases computation time and storage.
Just a decade ago the above memory requirements were large
for a standard desktop computer with a few megabytes memory. This may
explain why this fairly straightforward method has not been proposed before.

\begin{table}[tb]
    \caption[Rejection rate and number of tiles for a uni-modal PDF.]
    {\label{tab:unimodal} 
      Number of tiles, rejection rate and evaluation rate for the uni-modal PDF
      shown in Fig.~\ref{fig:asymmetric}, but with a 
      larger cutoff at $x=\pm 64$. Refinement level 5 is shown in
      Fig.~\ref{fig:asymmetric}. The  memory needed to store $49\,685$
      tiles (refinement level 10) is ca.\ 0.4~megabytes. The evaluation rate
      tells how often $f(x)$ must be evaluated per non-uniform random number.}

\begin{center}
\begin{tabular}{Sr Sr Sl Sl}
  \hline\hline
  \parbox{10ex}{Refinement\\ level $r$} &  \parbox{11ex}{Number of\\ tiles $N$} & \parbox{10ex}{Rejection rate $R$} & \parbox{10ex}{Evaluation rate $E$} \\
  \hline
  1 &          1 &        0.813   &   1      \\
  2 &          3 &         0.750  &    1      \\
  3 &          8 &         0.627   &   1      \\
  4 &          24 &         0.502 &     0.910      \\
  5 &          70 &         0.317 &     0.650      \\
  6 &         238 &         0.196   &   0.390      \\
  7 &         857 &         0.108  &    0.220      \\
  8 &    $3\,246$ &         0.058  &    0.110      \\
  9 &   $12\,609$ &         0.029  &    0.058      \\
 10 &   $49\,685$ &         0.015  &    0.029      \\
 11 &  $197\,233$ &         0.007  &    0.013      \\
 12 &  $785\,936$ &         0.002  &    0.005\\
  \hline\hline
\end{tabular}
\end{center}
\end{table}

\begin{table}[tb]
    \caption[Rejection rate and number of tiles for a multi-modal PDF.]{\label{tab:multitable}
      Statistics for the bimodal PDF shown in Fig.~\ref{fig:bimodal},
      where refinement level~6 is plotted.
      The  memory needed to store $151\,068$  tiles for refinement level 11 is ca.\ 1.2~megabytes.
}

\begin{center}
\begin{tabular}{Sr Sr Sl Sl}
  \hline \hline
 \parbox{10ex}{Refinement\\ level $r$} &  \parbox{11ex}{Number of\\ tiles $N$}  & \parbox{10ex}{Rejection rate $R$} & \parbox{10ex}{Evaluation rate $E$} \\
  \hline
  1 &          1     &    0.858   &        1     \\
  2  &         4     &    0.858   &        1     \\
  3  &         9     &    0.747   &        1     \\
  4  &        23     &    0.605   &   0.956     \\
  5  &        70     &    0.481   &   0.871     \\
  6  &       213     &    0.317   &   0.582     \\
  7  &       718     &    0.189   &   0.356     \\
  8  &  $2\,602$     &    0.106   &   0.195     \\
  9  &  $9\,859$     &    0.056   &   0.104     \\
  10 & $38\,324$     &    0.029   &   0.053     \\
  11& $151\,068$     &    0.014   &   0.025     \\
  12 &$599\,819$     &    0.007   &   0.011\\
  \hline\hline
\end{tabular}
\end{center}
\end{table}

\subsection{Speed of random variate production}

With the SHR3 uniform RNG~\cite{Marsaglia2000} on
the above mentioned configuration our method produces 2.6 million non-uniform random
numbers per second \emph{independently} of all tested PDFs.  In the
following we discuss a few pitfalls of speed measurement and code
execution, and we compare to other methods. The benchmarks 
refer to methods and implementations that appear most similar or 
useful in judging the tiling method. In any case, the comparisons
cannot be entirely fair since each method has different specialities.

In rejection methods the speed of random variate production is
arbitrarily independent of the PDF and its representation, whether
by data points or a closed formula. The speed depends only
on the properties of the comparison and squeeze functions.
In all our tested examples with tabular data or simple explicit
density functions the evaluations representing $f(x)$ are negligible
at a rejection rate below 0.02. Since interpolation or evaluation of
density functions is not the topic of this paper, we only give as a
rule of thumb that evaluations for 1\% of the produced random
numbers is sufficiently low for almost all practically relevant
densities.  The production of one random variate with the desired
distribution requires at least two uniform random variates as in most
methods. Recently a method was published that can provide non-uniform
variates with $1+s,\ s\in[0,1],$ uniform variates where $s$ can be made arbitrarily
small~\cite{Leydold2003a}.  However, it turns out that in almost all
applications the generation of uniform random numbers is not the
major sink of computer time. It is up to the scientist to evaluate
the trade-off between a few percent gain in overall speed and quality of
the obtained variates.  The use of less than two uniform random variates per
non-uniform variates in the context of a rejection technique but also
the importance of uniform random number quality, in particular in the
Ziggurat implementation by Marsaglia and Tsang, is commented in
Refs.~\cite{Brent2004,Doornik2005,Leong2005,Pang2001}.  
Some constructive remarks on the Ziggurat implementation in Ref.~\cite{Marsaglia2000}
can be found in Ref.~\cite{Nadler2006}.

We chose as one of the benchmarks the symmetric L\'evy $\alpha$-stable distribution.
It is a generalization of the Gaussian distribution, that is recovered for $\alpha=2$;
see Appendix.
The transformation method by Chambers et al.~\cite{Chambers1976} is the
contemporary method of choice. As opposed to other published methods
it has no accuracy deficiency, it does not truncate the support
and is sufficiently fast for most applications.
Moreover, it is applicable to asymmetric L\'evy $\alpha$-stable deviates too.
We use an implementation in C++ for the purpose of
this comparison. It is about 3 times faster than our method on the
above mentioned test configuration.

We also compared to the most efficient implementation of the Ziggurat
method~\cite{Marsaglia2000} for $\exp(-x)$ and $\exp(-x^2)$
distributed variates. This implementation is considered the fastest
for these two distributions.  The exponential and normal
densities could be wired into the code exploiting their mathematical
properties and using inline coding.  In the limit of a negligible
rejection rate, this Ziggurat implementation could produce 232 million
variates per second. This means one variate per 10 CPU clock cycles!
It is important to note that this number could only be achieved if
executed alone without any other code, for example within a Monte Carlo application.
This speed may be surprising at first sight since the rejection principle is
quite similar to the tiling method. Actually there are profound differences.
First and most obviously, the number of tiles is not a power of two. Choosing
randomly between exactly $2^8$ or $2^7$ objects is faster if one uses
8 bits of the 32 bit XOR shift RNG as in
Ref.~\cite{Marsaglia2000}. Secondly, it is stated self-evidently in
Ref.~\cite{Marsaglia2000} that small code is important. This purely
technical issue is hardly ever explained in the literature on random
numbers despite being highly technical on several
occasions. Numerical literature~\cite[Chap. 7]{Press2007} finally
picks up this issue and also more recently in Ref.~\cite{Rubin2006}, 
but only briefly say \emph{why}
small code is important. We outline the situation.

CPUs use hierarchical memory to speed up computation. The access to
the internal cache memory is magnitudes faster than to the external main
memory. However, the code and data fitting into this cache is not
the only condition for faster execution. An algorithm hard-wired in the
CPU transfers repeatedly and frequently used sections of memory into the
cache and also considers the size and distribution of the data over
the memory banks. A good implementation (and compiler) therefore tries
to minimize cache misses by arranging data of subsequent memory
accesses into the same cache line. The latter are sequences of bytes
transferred into the cache with each memory access.  This statistics
is disrupted by cache misses that are also provoked by a process
switch of the operating system at built-in time intervals or other
events.  Small code might therefore end up in the cache for a
significant time.  Very large code that accesses its data in random
fashion as it is the case in the sampling with tiles will not be able to
exploit properly cache memory.  We can therefore say that the
execution of code is subject to decisive factors of hardware,
compilation and operating system that can usually not be controlled
entirely.  It is also known that CPU-specific compilers are able to
produce code that can be several times faster than a more generic
compiler.

On our typical configuration of operating system and compiler the
execution of the Ziggurat code~\cite{Marsaglia2000} is the fastest by
far. The speed factor of ca.\ 100 to our code is in fact consistent to
the latency of low-level memory as compared to second-level cache of
contemporary hardware. This speed difference is leveled out
considerably if the code and
tables of the Ziggurat implementation is forced to leave the cache by
executing some arbitrary and larger code alternatingly with calls to
the Ziggurat generator.  This measure creates a more realistic use case
and reduces the execution speed of the Ziggurat code by a factor of ca.\ 50. 
A more rigorous analysis
of code and hardware interplay would require the \emph{exact} reproduction
of the original test environment which is not readily available anymore.

Finally we make a few more technical remarks and comparisons.
The implementation of the tiling method is only moderately optimized,
but completely portable and uses throughout Standard Template Library
arrays. The period of the XOR shift RNG is
considered short with 32 bit arithmetic but modification to higher models
is possible. Following the results in Refs.~\cite{Doornik2005,Nadler2006,Pang2001}
on quality, resolution and portability we recommend a slower and also portable
uniform RNG.  Refs~\cite{Brent2004,Doornik2005,Leong2005,Nadler2006,Pang2001} also
comment other problems of the XOR shift RNG in conjunction with the Ziggurat method. 
The Ziggurat method requires for the decision whether to evaluate the
density function one coordinate comparison for each attempt to draw a
non-uniform number.  Our method requires one table look-up only. But
this advantage is not enough to compensate the disadvantage of a large
table and resulting slow memory access.

For accelerated production of
random variates to make sense, their part must take up a significant proportion of
the overall CPU time.  But there is hardly anything do-able
within the order of 10 clock cycles. Moreover, fast production of variates imply
that enormous amounts are required. This poses very high demands on their quality.
The findings above as well as the critical publications on the Ziggurat implementation
Ref.~\cite{Marsaglia2000} encourage  to
analyse the appropriateness of extremely fast but medium quality variates. A detailed
analysis of this issue can be found in Ref.~\cite{Thomas2007b}.

\subsection{Speed measurements of the setup}

The setup part in our implementation is not speed-optimized but turned
out to be sufficiently fast for the production of ca. one million
variates and above.  This includes the extreme examples with more than
one mode and a very large support.  To provide a meaningful time
measurement for the setup we subtract the cumulative time for the
evaluations of $f(x)$.  For the presented examples we used a standard
polynomial interpolation with 7 data points.  The setup for a typical
uni-modal PDF (Gaussian or L\'evy, the latter with sufficiently wide
support) with $2^{15}$ data points takes ca.\ 0.2~seconds plus
cumulative 2.1~seconds for all evaluations of $f(x)$.  The calculation
time of the L\'evy PDF via fast Fourier transform for $2^{15}$ points is
negligible with only 0.2~seconds.  Thus, as a rule of thumb, the overall total speed of
the setup depends almost entirely on the number of evaluations of
$f(x)$. With a constant number of
data points the total speed of the setup increases noticeably only for
very unusual multi-modal PDFs with many sharp peaks and long tails.

The setup of the Ziggurat for general symmetric, strictly decreasing,
non-analytic and safely truncatable PDFs was attempted in
Ref.~\cite{Leccardi2005}. This setup, our C++ version of the Matlab code from
Ref.~\cite{Leccardi2005} as well as our own generalized iterative
C++ code along the original Ziggurat setup formula~\cite{Marsaglia2000} is sensitive
and computationally expensive. 
For example, a numerical error in the flat regions of the tail or in the
inversion of the PDF can cause a disturbance  which often causes a breakdown of the
procedure.  Precautions to mend this are possible but complicate the
code further and do not guarantee unattended functionality. The empirical parameters
needed for the setup of the Ziggurat method are an additional difficulty for
making the method truly automatic. The setup time depends strongly on the given data
and the above mentioned empirical parameters, and is at least one order of
magnitude slower than the tiling.

\section{Discussion and conclusion}\label{sec:conclusion}

We presented a fast method for automatic generation of random variates with
arbitrary probability density functions independent of symmetry, number of modes,
and discontinuities. The only prerequisites are pointwise computability
and finite support. We also explained that the most general thinkable
or universal method will require no less but also no more than these
two requirements. In the introductory overview on some representative
methods it is shown that many less powerful methods exist that
truncate the infinite support for analytic density functions with only
one mode.  The accuracy of our method is exact up to
the computation of the probability density function and meets 
any numerical demand which includes density functions with poles or
cusps without additional attention.

The generation of one non-uniform random variate requires 
only one random integer, one random uniform real, two additions, one
multiplication and one table look-up (no float comparison) most of the
time.  This is close to the minimum of principally required operations,
so that additional speed can only come from hardware exploitation or
specialized methods. Even for complicated density functions the
memory requirements are suitable for any contemporary desktop
computer.

These properties are not available in other methods of this kind.
We can  extend the wish list from Sec.~\ref{sec:tiling}
to a random number generator by additional items:
\begin{enumerate}
\item[8.] No need for a priori knowledge about the location of
  \emph{any number} of modes or discontinuities within the density
  function.
\item[9.] Only pointwise computability and representability of the
  density function is necessary.
\item[10.] Fast setup time and fast generation of random variates.
\item[11.] The discretization and therefore sampling efficiency is
  \emph{asymptotically exact} and can be pre-imposed.
\end{enumerate}

An extension to the multivariate case is simple in principle. It
means to substitute a two-dimensional tile with a cube or hypercube.
The required storage for data however increases with a power of the dimension.


\section{Appendix: L\'evy density function}
\label{sec:levy}

The generation of the pointwise density function of the L\'evy $\alpha$-stable 
distribution and numbers is a technicality that we report for completeness.
Any density would suffice for purpose of proving correctness numerically
and there are certainly more cumbersome densities in statistical computing but
we prefer also to have inversion methods at hand to produce 
random numbers. We choose the formula by Chambers et al., 1976~\cite{Chambers1976}:
\begin{equation} \label{eqn:} 
  \xi = \gamma\left(\frac{-\log
      u_1\cos\phi}{\cos((1-\alpha)\phi)}\right)^{1- \frac{1}{\alpha}}
  \frac{\sin(\alpha\phi)}{\cos\phi},
\end{equation}
where $\phi = \pi(u_2-1/2)$; $u_1,\ u_2 \in (0,1)$ are uniformly
distributed random numbers, $\gamma$ is the scaling parameter, and
$\xi$ is a L\'evy distributed random number. This expression, however,
requires different representations for certain ranges of $\alpha$ to
reduce numerical error.  Direct coding is possible but not
recommended.
 
The pointwise calculation of the PDF is more difficult. It requires
the calculation of a very slowly converging
integral~\cite{Nolan1997,Nolan1999}.  In the symmetric case the L\'evy
distribution can be defined by
\begin{equation} \label{eqn:f_Levy}
L(z; \alpha, \gamma)=\frac{1}{\pi}\int_0^\infty \exp({-(\gamma q)^\alpha})\:\cos(qz)\: dq\; .
\end{equation}
This is a parade example of a non-analytic density function.  It does
not possess any moments other than the first for $\alpha\in(1,2]$ due
to the divergence of the respective integrals except for $\alpha=2$
which is the Gaussian limit.  Values for $L(z)$,
Eq.~(\ref{eqn:f_Levy}), can be computed
directly~\cite{Nolan1997,Nolan1999} or by carrying out the Fourier
Transform~\cite{Mittnik1999}.


\bibliographystyle{amsplain}
\bibliography{journals,paper}

\end{document}